\documentclass[lettersize,journal]{IEEEtran}
\usepackage{amsmath,amsfonts}
\usepackage{algorithmic}
\usepackage{algorithm}
\usepackage{array}
\usepackage{textcomp}
\usepackage{stfloats}
\usepackage{url}
\usepackage{verbatim}
\usepackage{graphicx}
\usepackage{cite}
\usepackage{amssymb}
\usepackage{subfigure}
\usepackage{multirow}
\usepackage{multicol}
\usepackage[switch]{lineno}
\usepackage{color}

\usepackage{booktabs}
\begin{document}

\title{Unveiling Causalities in SAR ATR: A Causal Interventional Approach for Limited Data}

\author{Chenwei Wang,~\IEEEmembership{Member,~IEEE,}
        Xin Chen,
        You Qin,
        Siyi Luo,\\
        Yulin Huang,~\IEEEmembership{Senior Member,~IEEE,}
        Jifang Pei,~\IEEEmembership{Member,~IEEE,} 
        and Jianyu Yang,~\IEEEmembership{Member,~IEEE}
\thanks{This work was supported by the Sichuan Science and Technology Program under Grant 2023NSFSC1970.(\emph{Corresponding author: Yulin Huang.)}}
\thanks{The authors are with the Department of Electrical Engineering, University of Electronic Science and Technology of China, Chengdu 611731, China (e-mail: peijfstudy@126.com; dbw181101@163.com).}}

\maketitle

\begin{abstract}
Synthetic aperture radar automatic target recognition (SAR ATR) methods fall short with limited training data.
In this letter, we propose a causal interventional ATR method (CIATR) to formulate the problem of limited SAR data which helps us uncover the ever-elusive causalities among the key factors in ATR, and thus pursue the desired causal effect without changing the imaging conditions.
A structural causal model (SCM) is comprised using causal inference to help understand how imaging conditions acts as a confounder introducing spurious correlation when SAR data is limited. 
This spurious correlation among SAR images and the predicted classes can be fundamentally tackled with the conventional backdoor adjustments. 
An effective implement of backdoor adjustments is proposed by firstly using data augmentation with spatial-frequency domain hybrid transformation to estimate the potential effect of varying imaging conditions on SAR images.
Then, a feature discrimination approach with hybrid similarity measurement is introduced to measure and mitigate the structural and vector angle impacts of varying imaging conditions on the extracted features from SAR images.
Thus, our CIATR can pursue the true causality between SAR images and the corresponding classes even with limited SAR data. 
Experiments and comparisons conducted on the moving and stationary target acquisition and recognition (MSTAR) and OpenSARship datasets have shown the effectiveness of our method with limited SAR data.
\end{abstract}

\begin{IEEEkeywords}
SAR, ATR, Causal Graph, Interventional Training
\end{IEEEkeywords}

\section{Introduction}
\IEEEPARstart{S}{ynthetic} aperture radar (SAR) is a flexible remote sensing technology, useful in multiple civil and military contexts, offering high-resolution imaging regardless of time or weather \cite{intro1,wang2022sar,wang2021multiview}. Its key application, automatic target recognition (ATR), has evolved over the past fifty years \cite{intro2,wang2020deep,wang2019parking}. Notably, the last decade has seen considerable improvements in ATR's performance, largely driven by advancements in deep learning technology \cite{comparison1,comparison2,isprs2,li2023panoptic,liang2023efficient, wang2022recognition,wang2023entropy,wang2021deep}.

While current deep learning-based SAR ATR methods exhibit encouraging performance, they inherently rely on extensive information drawn from a large collection of SAR images. 
However, the process of generating a substantial volume of SAR images, coupled with the requirement for accurate labelling, is both resource-intensive and time-consuming \cite{comparison3,wang2023sar}. 
As a result, there is often a shortage of information available for supervised training, which ultimately impacts the effectiveness of these methods. 
This issue underscores a critical disconnect between the theoretical design of ATR methods and their practical applications, with existing methods often falling short when deployed in real-world contexts \cite{compared4,wang2022global,wang2020multi}. This problem, termed as SAR ATR with limited training data, has recently become a focus in research \cite{isprs1,compared2,compared3,wang2022semi}.

\begin{figure}
\begin{center}
\subfigure[Causal Graph for ATR with limited SAR data]{\includegraphics[width=0.15\textwidth]{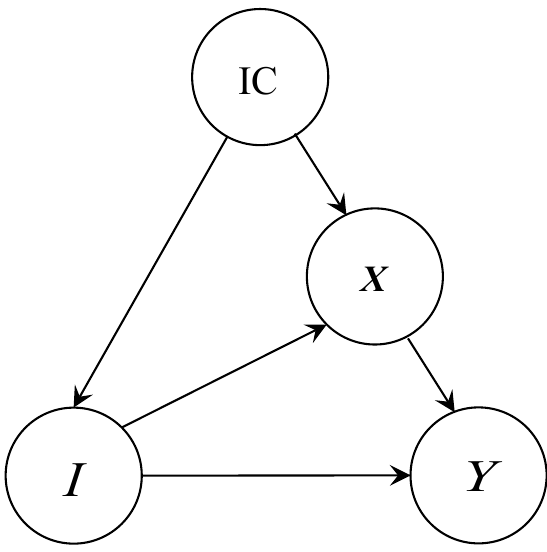}}
\quad
\subfigure[Sensitiveness of SAR images to Imaging Conditions]{\includegraphics[width=0.15\textwidth]{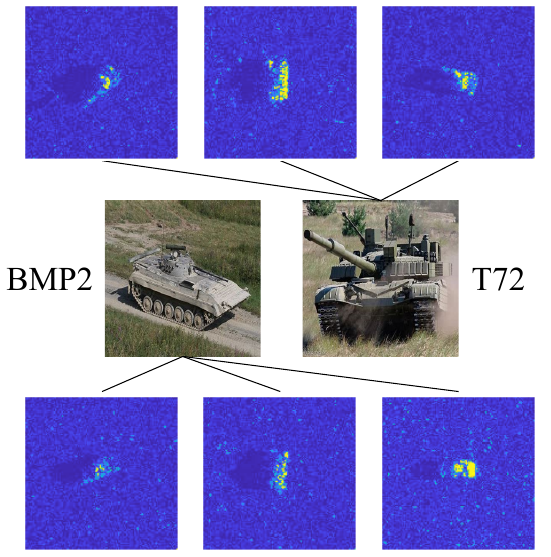}} 
\subfigure[Our CIATR]{\includegraphics[width=0.15\textwidth]{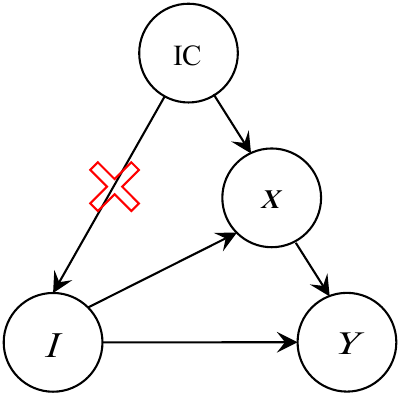}} 
\end{center}
\caption{Causal graph for ATR with limited SAR data. (a) $I$ is SAR images, $IC$ is the imaging conditions, $X$ is the features affected by $IC$, and $Y$ is the predicted classes. (b) Even only the azimuth angle is changing, the scattering characteristics of SAR images varies. (c) is our CIATR as a solution of the decounding training for ATR with limited SAR data.}
\label{problemstate}
\end{figure}

The primary challenge in ATR with limited training data is the weak performance caused by the sensitivity of SAR images to imaging conditions, as shown in the causal graph in Fig. \ref{problemstate} (a).
The causal graph is constructed based on the assumption of the causalities among the SAR images $I$, the imaging conditions $IC$, the extracted feature $X$, and the classification $Y$.
When the imaging conditions IC are changing, the inner-class SAR images have obvious variance of scattering characteristics. Besides, it is inevitable that the features X contains some feature IF effected by IC.
As shown in Fig. \ref{problemstate} (b), even only a change in the azimuth angle causes inner-class SAR images to display distinct scattering characteristics.
In the process of recognition, the ATR method aims to model $P(Y|X)$, but the imaging conditions $IC$ act as a confounder that is the common cause of the features via $IC \to X$ and the classification Y via $M \to IF \to Y$. 
As a results, the imaging conditions introduce spurious correlation in the process of modeling $P(Y|X)$, thus leading to the weak performance of ATR method with limited SAR data. 

Therefore, in this letter, we propose a causal interventional ATR method (CIATR) with limited SAR data that not only fundamentally analysis the role of imaging conditions in the recognition, but also provides a principled solution to improve the recognition performance. Specifically, our contributions are summarized as follows.

1) Section 2.1 introduces a Structural Causal Model (SCM) that elucidates why imaging conditions, while negligible with ample SAR data, act as a confounder introducing spurious correlation into the ATR model when data is limited.

2) Section 2.2 then outlines specific effective implementations using backdoor adjustment \cite{glymour2016causal}, which mainly consist two steps: data augmentation with spatial-frequency domain hybrid transformation, and the feature discrimination with hybrid similarity measurement.

3) Thanks to the causal intervention, the CIATR achieves state-of-the-art recognition performances on MSTAR and OpenSARship data set with different numbers of training samples. The ablation experiments have validated the effectiveness of the CIATR.

The rest of this letter is organized below. Section \uppercase\expandafter{\romannumeral2} presents the causal graph and solution for ATR with limited SAR data. Section \uppercase\expandafter{\romannumeral3} verifies the effectiveness of the proposed method with experiments, and Section \uppercase\expandafter{\romannumeral4} gives our conclusion.

\begin{figure*}[tb]
\centering
\includegraphics[width=0.9\textwidth]{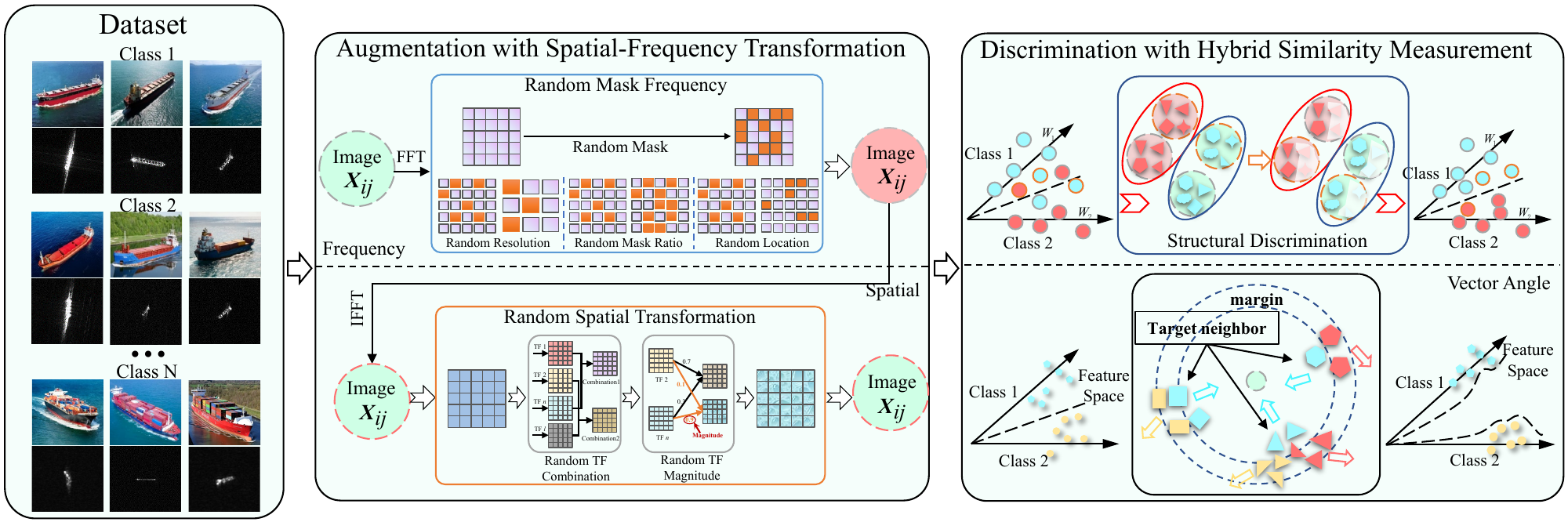}
\caption{Specific implement of backdoor adjustments in our CIATR.}
\label{implement}
\end{figure*}

\section{Causal Interventional ATR Method}
This section starts with the introduction of a Structural Causal Model (SCM), detailing the cause-and-effect relationships among variables in SAR ATR. 
A core solution is proposed to mitigate the false correlations arising due to imaging conditions, thereby enhancing recognition performance even when SAR data is scarce.

\subsection{Structural Causal Model}
To systematically analyze ATR with limited SAR data, the SCM is a directed acyclic graph that elucidates the influence of variables of interest ${\bf{I}}$, ${\bf{X}}$, $\rm{IC}$ on the ATR model's recognition results, $Y$. Each arrow represents a causal relationship between two nodes. The fundamental reasoning behind SCM is detailed below.

${\bf{I}} \to Y$. The ATR modal aims to finish the precise recognition $Y$ condition on ${\bf{I}}$, $P(Y|{\bf{I}})$. There are two paths to determine $Y$ by ${\bf{I}}$: 1) ${\bf{I}} to Y$ is the direct path which means ${\bf{I}}$ has a direct effect on $Y$. 2) ${\bf{I}} \to {\bf{X}}_v \to Y$ is the mediation path which means that the features ${\bf{X}}_v$ extracted from $\bf{I}$ play the mediator role in the recognition process. 

${\rm{IC}} \to {\bf{I}} \to Y$. The sensitivity of SAR images ${\bf{I}}$ to the imaging conditions ${\rm{IC}}$ leads to the scattering characteristic varying when ${\rm{IC}}$ is changing. The imaging conditions contain many aspects, for example, the azimuth angle and the parameters of the imaging platform. Even if one factor of imaging conditions changes, i.e., the azimuth angle, the scattering characteristic of SAR image varies obviously. As a result, the imaging conditions ${\rm{IC}}$ affect the scattering characteristics of the entire SAR image ${\bf{I}}$, thereby influencing the classification $Y$.

${\rm{IC}} \to {\bf{X}}_v \leftarrow {\bf{I}}$. The features ${\bf{X}}_v$ are denoted as the low-dimensional representations of SAR image $\bf{I}$. 1) ${\rm{IC}} \to {\bf{X}}_v$. The features ${\bf{X}}_v$ contains the features from the targets and the background, the imaging conditions ${\rm{IC}}$ have an obvious effect on the background, like cluster and shadow regions. Thus, the imaging conditions ${\rm{IC}}$ and the SAR image $\bf{I}$ have the causality for the feature ${\bf{X}}_v$. When modeling the recognition $P(Y|{\bf{I}})$, the information of $\rm{IC}$ affects not only the SAR images but also the extracted features ${\bf{X}}_v$. 

An ideal ATR model with limited SAR data should rely on the true causality between $\bf{I}$ and $Y$ to achieve precise recognition. However, as mentioned above, the conventional modeling $P(Y|\bf{I})$ fails to be ideal, because the likelihood of $Y$ given $\bf{I}$ is not only due to ${\bf{I}} \to Y$ and ${\bf{I}} \to {\bf{X}}_v \to Y$, but also the spurious correlation introduced by $\rm{IC}$ via ${\bf{D}} \to {\bf{I}}$ and ${\bf{D}} \to {\bf{X}}_v \to Y$.

Therefore, to obtain an ideal ATR model with limited SAR data, it is necessary to pursue the true causality between the $I$ and $Y$ without the spurious correlation introduced by $\rm{IC}$.
Fortunately, the backdoor adjustment can be used to implement the causal intervention $P(Y|do({\bf{I}}))$ to mitigate the spurious correlation introduced by $IC$:
\begin{align}
P(Y|do({\bf{I}})) &= \sum_{{\bf{X}}_v}\sum_{\rm{IC}} P(Y|{\bf{I}}, {\bf{X}}_v, {\rm{IC}})P(X_v | {\bf{I}}, {\rm{IC}})P({\rm{IC}})\\
&= \sum_{{\bf{X}}_v}\sum_{\rm{IC}} P(Y| {\bf{I}}, {\bf{X}}_v)P(X_v | {\bf{I}}, {\rm{IC}})P({\rm{IC}})\\
&= \sum_{\rm{IC}} P(Y| {\bf{I}}, {\rm{IC}},{\bf{X}}_v=g(\rm{I},\rm{IC}))P({\rm{IC}})
\end{align}
Due to the rule 1 of do-Calculus \cite{pearl2012calculus}, $\mathbf{D}$ does not affect $Y$ directly, so $P(Y| \bf{I}, C, D)$ in Eq. (1) can be replaced by $P(Y| {\bf{I}},{\bf{X}}_v)$, yielding Eq. (2). Eq. (3) is because, in our SCM, $X_v$ takes a deterministic value given by function $g(I,IC)$. These equations above means that if the imaging conditions are observable, it is possible to employ the physical intervention by stratifying $\rm{IC}$ to mitigate the spurious correlation introduced by $\rm{IC}$. Stratifying $\rm{IC}$ means producing an integrated set of ${\bf{X}}_v)$ using every value of $IC$ for any given SAR image $i$. Therefore, in the process of modeling $P(Y|do({\bf{I}}))$, eliminating the influence of $\rm{IC}$ and ${\bf{X}}_v)$ can achieve Eq. (2).

In the following sections, based on the solution above, we present the specific effectiveness implementation to improve recognition performance with limited SAR data.

\subsection{Interventional Augmentation and Discrimination}

Fig. \ref{implement} illustrates our approach. Initially, we augment data using transformations in image and frequency domains, simulating varied imaging conditions, a process akin to $\rm{IC}$ accumulation in Eq. (2) \cite{pearl2012calculus}. Next, we apply a hybrid similarity measure for feature discrimination, calculating the effect of different conditions on a SAR image. Using the invariant risk minimization (IRM) concept, we provide a loss $L_d$ to enable ${\bf{X}}_v$ accumulation in Eq. (2). By minimizing $L_d$, our CIATR models achieve precise recognition with limited SAR data.
The details of the data augmentation with spatial-frequency domain hybrid transformation and the feature discrimination with hybrid similarity measurement are as follows.

Given the limited SAR training set ${\bf{D}}^{tr}=\left \{{\bf{D}}_1^{tr},{\bf{D}}_2^{tr},..,{\bf{D}}_C^{tr} \right \}$, where ${\bf{D}}_i^{tr}=\left \{{\bf{x}}_{i1},…,{\bf{x}}_{in}\right \}$ represents the training samples of the $i$th class, and ${\bf{x}}_{ij} \in \mathbb{R}^{h \times w}$ is the $j$th sample in ${\bf{D}}_i^{tr}$, with $n$ being the sample number of each class. 

As illustrated in Fig. \ref{implement}, the spatial-frequency transformation augmentation comprises two aspects: random frequency mask and spatial transformation. Firstly, each image in ${\bf{D}}^{tr}$, such as ${\bf{x}}_{ij}$, undergoes a fast Fourier transform (FFT) to derive its spectrum. We then apply a random mask to maximize potential imaging condition estimations. The random resolution, $rm_{re}$, denotes the smallest resolution unit in the augmentation, implying ${\bf{x}}_{ij}$ is split into $h \times w/({rm}_{re})^2$ patches. The mask ratio, ${rm}_{ra}$, indicates the number of zeroed patches, while the location, $rm_l$, represents the index of these patches.
Thus, the process of random frequency mask can be presented as ${\bf{x}}^f_{ij}= RFM(fft({\bf{x}}_{ij}),{rm}_{re},{rm}_{ra},{rm}_l)$, where ${\bf{x}}^f_{ij}$ is the masked version of ${\bf{x}}_{ij}$, $RFM(\cdot)$ is the operation of random frequency mask, and $fft(\cdot)$ is the fast Fourier transform. 

Then the inverse FFT is employed to convert ${\bf{x}}^f_{ij}$ into the spatial domain, and the random spatial transformation with the complete transformation set $TF=\{{tf}_1,…,{tf}_Q\}$, where ${tf}_i$ is the $i$th transformation, $Q$ is the class number of all the transformation. There are also two random variances. 
The random transformation combination $q$ is obtained by randomly sampling from $TF$, and $M_q$ is a random value following a normal distribution.
Thus, the process of random spatial transformation can be presented as ${\bf{x}}^s_{ij}= RST(ifft({\bf{x}}^f_{ij}),q, M_q)$, where ${\bf{x}}^s_{ij}$ is the final version of ${\bf{x}}_{ij}$ after the augmentation, $RST(\cdot)$ is the operation of random spatial transformation, and $ifft(\cdot)$ is the inverse FFT.

Through the process outlined above, each sample in ${\bf{D}}^{tr}$ is augmented with an additional random version to estimate the potential impact of various imaging conditions on SAR samples. It's worth noting that we've introduced multiple randomness factors to estimate as many imaging conditions as possible, ensuring the comprehensiveness of stratifying $\rm{IC}$.

The limited SAR training set can be presented as ${\bf{D}}_a^{tr}=\{{\bf{D}}_1^{tr},{\bf{D}}_2^{tr},..,{\bf{D}}_C^{tr} \}, {\bf{D}}_i^{tr}=\{{\bf{x}}_{i1},{\bf{x}}^s_{i1},…,{\bf{x}}_{in},{\bf{x}}^s_{in}\}$. 
Then, a feature discrimination method employs a structural measurement to enable the ATR model to capture effective local features. Concurrently, it uses a vector angle measurement to enhance the discriminability of the extracted features. The hybrid measurement aims to fulfill the summation over ${\bf{X}}_v$ in Eq. (2).
The process of feature discrimination consists of two parts: hybrid measurement and loss calculation, as shown in Fig. \ref{implement}. 
For every pair of samples in ${\bf{D}}^{tr}$, we calculate their hybrid measurement:
\begin{equation}
hm({\bf{x}}_{ij},{\bf{x}}_{nm})=stm({\bf{x}}_{ij},{\bf{x}}_{nm})+vam({\bf{x}}_{ij},{\bf{x}}_{nm})
\end{equation}
where $hm(x,y)$ is the hybrid measurement between $x$ and $y$, $stm(x, y)$ is the structural measurement between $x$ and $y$, and $vam(x, y)$ is the vector angle measurement. The structural similarity index measure (SSIM) is employed as $stm(\cdot,\cdot)$, and the cosine similarity is employed as $vam(\cdot,\cdot)$.

Then, if ${\bf{x}}_{ij}$ and ${\bf{x}}_{nm}$ belong to the same class, i.e., $j=m$, $hm({\bf{x}}_{ij},{\bf{x}}_{nm})$ should be as large as possible. Conversely, if ${\bf{x}}_{ij}$ and ${\bf{x}}_{nm}$ do not belong to the same class, i.e., $j!=m$, $hm({\bf{x}}_{ij},{\bf{x}}_{nm})$ should be as small as possible.
Thus, the discrimination loss $L_d$ can be calculated based on the triplet loss.
Besides, the cross-entropy loss is employed as the basic recognition loss.
The final loss is the summation of $L_{ce}$ and $L_d$.

Our CIATR first proposed an SCM to analyze the reason for weak performance of ATR with limited SAR data and provides a causal solution to pursue the true causality between the $\bf{I}$ and $Y$ without the spurious correlation introduced by $\rm{IC}$.

\section{Experiments}
In this section, we assess the effectiveness of our method using benchmark SAR image datasets, OpenSARship and MSTAR.

\subsection{Dataset}
The OpenSARship dataset, gathered from 41 diverse Sentinel-1 images, facilitates the development of sophisticated ship detection and classification algorithms for challenging environments. This dataset comprises 11346 slices from 17 SAR ship types, integrated with reliable AIS information. Experiments utilize the GRD data, featuring a $2.0{\rm m} \times 1.5{\rm m}$ resolution and a $10{\rm m} \times 10{\rm m}$ pixel size in Sentinel-1 IW mode. Ship dimensions span from 92m to 399m in length and 6m to 65m in width.

The MSTAR dataset, a SAR ATR performance assessment standard, was launched by the Defense Advanced Research Project Agency and the Air Force Research Laboratory. Acquired via Sandia National Laboratory's STARLOS sensor, it includes X-band SAR images with 1-ft resolution across a 0° to 360° range. 

\begin{table}[tb]
\renewcommand{\arraystretch}{1.2}
\setlength\tabcolsep{16.4pt}
\centering
\footnotesize 
\caption{Image Number of Different Targets of OpenSARship Dataset}
\label{ttnumOPEN}
\begin{tabular}{c|c|c|c}
\toprule\toprule
Class          & Training & Testing & Total  \\ \midrule
Bulk Carrier   & 300      & 374     & 674  \\ 
Container Ship & 300      & 710     & 1010 \\ 
Tanks          & 300      & 253     & 553  \\
Cargo   & 300      & 557     & 857  \\ 
Fishing & 300      & 121     & 421 \\ 
General Cargo          & 300      & 165     & 465  \\ \bottomrule\bottomrule
\end{tabular}
\end{table}

\subsection{Recognition Performances and Comparisons under OpenSARship and MSTAR}
In this section, the experimetns under OpenSARship and MSTAR dataset is run and presented.

\subsubsection{Recognition Performances and Comparisons of 3 and 6 classes under OpenSARship}
The OpenSARShip dataset comprises several ship categories, representing 90\% of the international shipping market, the most common and significant ships \cite{compared3}. As per \cite{open4}, experiments were conducted considering different numbers of classes: 3 and an extended set of 6. The 3-class experiment incorporates bulk carriers, container ships, and tanks, while the extended 6-class set also includes cargo ships, fishing vessels, and general cargo (Table \ref{ttnumOPEN}).

\begin{table}[tb]
\renewcommand{\arraystretch}{1.2}
\setlength\tabcolsep{3.1pt}
\centering
\footnotesize 
\caption{Recognition Performance (\%) of 3 Classes under Different Training Data in OpenSARship Dataset}
\label{performanceOPEN}
\begin{tabular}{c|ccccccccc}
\toprule \toprule
\multirow{2}{*}{Class} & \multicolumn{8}{c}{Training Number in   Each Class}                          \\ \cline{2-9} 
                       & 20    & 30    & 40    & 50   &60     & 70    & 80   & 100   \\ \hline
Bulk Carrier           & 71.16 & 65.89 & 58.74 & 68.42 & 67.16 & 74.11 & 71.79 & 78.74 \\
Container Ship         & 62.52 & 70.90 & 81.63 & 81.50 & 79.53 & 78.91 & 83.23 & 78.79 \\
Tanker                 & 85.88 & 86.72 & 80.79 & 81.07 & 87.29 & 83.33 & 87.01 & 88.42 \\ \midrule
Average                & 70.06 & 72.87 & 74.82 & 77.62 & 77.62 & 78.48 & 80.73 & 80.85 \\ \bottomrule\bottomrule
\end{tabular}
\end{table}

\begin{table}[tb]
\renewcommand{\arraystretch}{1.2}
\setlength\tabcolsep{3.1pt}
\centering
\footnotesize 
\caption{Recognition Performance (\%) of 6 Classes under Different Training Data in OpenSARship Dataset}
\label{6-performanceOPEN}
\begin{tabular}{c|cccccccc}
\toprule \toprule 
\multirow{2}{*}{Class} & \multicolumn{8}{c}{Training Number in   Each Class}                          \\ \cline{2-9} 
                       & 20    & 30    & 40    & 50   &60     & 70    & 80   & 100   \\ \hline
Bulk Carrier                               & 64.42 & 59.16 & 48.84 & 58.95 & 65.68 & 68.42 & 61.47 & 64.00 \\
Container Ship                             & 55.86 & 69.30 & 78.30 & 67.94 & 73.37 & 75.34 & 75.46 & 82.86 \\
Tanker                                     & 50.28 & 50.85 & 40.96 & 55.93 & 57.91 & 45.48 & 55.65 & 51.41 \\
Cargo                                      & 36.09 & 36.09 & 43.63 & 42.73 & 38.96 & 38.78 & 41.11 & 43.45 \\
Fishing                                    & 85.12 & 87.60 & 93.39 & 86.78 & 90.08 & 95.87 & 87.60 & 93.39 \\
General Cargo                              & 38.79 & 50.91 & 43.64 & 42.42 & 46.67 & 53.33 & 53.94 & 44.85 \\ \midrule
Average                                    & 52.56 & 56.95 & 57.99 & 58.07 & 61.01 & 61.10 & 61.42 & 63.91 \\ \bottomrule\bottomrule
\end{tabular}
\end{table}

Tables \ref{performanceOPEN} and \ref{6-performanceOPEN} illustrate our method's superiority in 3-class and 6-class SAR ship image recognition tasks with training samples per class between 20 and 100. The 3-class recognition rate rises from 70.06\% with 20 samples to 77.62\% with 50, demonstrating effective use of additional samples. In 6-class recognition, performance ascends from 52.55\% with 20 samples to 63.91\% with 100.

\renewcommand{\arraystretch}{2}
\begin{table}[]
\centering
\footnotesize
\setlength\tabcolsep{7.0pt}
\caption{Comparison of Performances (\%) of 3 classes under OpenSARShip (The number in parentheses is the number of the training samples for each method)}
\label{3comparisonOPEN}
\begin{tabular}{c|ccc}
\toprule \toprule 
\multirow{2}{*}{Methods} &
  \multicolumn{3}{c}{Number range of training images in each class} \\ \cline{2-4} 
 &
  \multicolumn{1}{c|}{1 to 50} &
  \multicolumn{1}{c|}{51 to 100} &
  101 to 338 \\ \hline
Supervised \cite{open4} &
  \multicolumn{1}{c|}{\begin{tabular}[c]{@{}c@{}}58.24 (20)\\ 62.09 (40)\end{tabular}} &
  \multicolumn{1}{c|}{65.63 (80)} &
  \begin{tabular}[c]{@{}c@{}}68.75 (120)\\ 70.83 (240)\end{tabular} \\ \hline
CNN\cite{compared2} &
  \multicolumn{1}{c|}{62.75 (50)} &
  \multicolumn{1}{c|}{68.52 (100)} &
  73.68 (200) \\ \hline
CNN+Matrix\cite{compared2} &
  \multicolumn{1}{c|}{72.86 (50)} &
  \multicolumn{1}{c|}{75.31 (100)} &
  77.22 (200) \\ \hline 
PFGFE-Net\cite{compared3} &
  \multicolumn{1}{c|}{-} &
  \multicolumn{1}{c|}{-} &
  79.84 (338) \\ \hline
MetaBoost\cite{compared4} &
  \multicolumn{1}{c|}{-} &
  \multicolumn{1}{c|}{-} &
  80.81 (338) \\ \hline
Proposed &
  \multicolumn{1}{c|}{\begin{tabular}[c]{@{}c@{}}70.06 (20)\\ 74.82 (40)\end{tabular}} &
  \multicolumn{1}{c|}{\begin{tabular}[c]{@{}c@{}}80.73 (80)\\ 80.85 (100)\end{tabular}} &
  85.40 (200) \\ \bottomrule \bottomrule
\end{tabular}
\end{table}

\begin{table}[h]
\renewcommand{\arraystretch}{1.2}
\setlength\tabcolsep{7.4pt}
\centering
\footnotesize
\caption{Original Image Number of Different Depressions}
\label{ttnumMSTAR}
\begin{tabular}{c|cc|cc}
\toprule \toprule
\multirow{2}{*}{Class} & \multicolumn{2}{c|}{Training}            & \multicolumn{2}{c}{Testing}             \\ \cline{2-5} 
                       & \multicolumn{1}{c|}{Number} & Depression & \multicolumn{1}{c|}{Number} & Depression \\ \hline
BMP2-9563 & \multicolumn{1}{c|}{233} & \multirow{9}{*}{$\text{17}{}^\circ$} & \multicolumn{1}{c|}{195} & \multirow{9}{*}{$\text{15}{}^\circ$} \\ \cline{1-2} \cline{4-4}
BRDM2-E71              & \multicolumn{1}{c|}{298}    &            & \multicolumn{1}{c|}{274}    &            \\ \cline{1-2} \cline{4-4}
BTR60-7532             & \multicolumn{1}{c|}{256}    &            & \multicolumn{1}{c|}{195}    &            \\ \cline{1-2} \cline{4-4}
BTR70-c71              & \multicolumn{1}{c|}{233}    &            & \multicolumn{1}{c|}{196}    &            \\ \cline{1-2} \cline{4-4}
D7-92                  & \multicolumn{1}{c|}{299}    &            & \multicolumn{1}{c|}{274}    &            \\ \cline{1-2} \cline{4-4}
2S1-b01                & \multicolumn{1}{c|}{299}    &            & \multicolumn{1}{c|}{274}    &            \\ \cline{1-2} \cline{4-4}
T62-A51                & \multicolumn{1}{c|}{299}    &            & \multicolumn{1}{c|}{273}    &            \\ \cline{1-2} \cline{4-4}
T72-132                & \multicolumn{1}{c|}{232}    &            & \multicolumn{1}{c|}{196}    &            \\ \cline{1-2} \cline{4-4}
ZIL131-E12             & \multicolumn{1}{c|}{299}    &            & \multicolumn{1}{c|}{274}    &            \\ \cline{1-2} \cline{4-4}
ZSU234-d08             & \multicolumn{1}{c|}{299}    &            & \multicolumn{1}{c|}{274}    &            \\ \bottomrule \bottomrule
\end{tabular}
\end{table}

In comparison with other methods (Table \ref{3comparisonOPEN}), our method excels. It yields rates of 70.06\% and 74.81\% when trained with 20 and 40 samples respectively, outperforming the Supervised method's 58.24\% with 20 samples. Furthermore, our method surpasses Supervised, CNN, and CNN+Matrix recognition rates with 80 samples. Thus, within the 1-50 samples band, our method outmatches state-of-the-art techniques.

In conclusion, the results under OpenSARship dataset illustrates the resilience and effectiveness of our method with limited samples, proving its capability to perform SAR ATR tasks under resource-constrained conditions.

\subsubsection{Recognition Performances and Comparisons under MSTAR}
In this subsection, we discuss the recognition performance of our CIATR and compare it with other algorithms under MSTAR dataset. In the K-shot setting, both our CIATR and other methods randomly select K images from each class in the MSTAR dataset for training in Table \ref{ttnumMSTAR}.

Table \ref{soc} shows the recognition results when the training samples are ranging from 5 to 100. Examining the recognition rate relative to the growth in sample size, we find that the rate increases significantly from 75.05\% to 86.47\% as samples double from 5 to 10. The rate continues to rise, reaching 96.70\% with 25 samples and 97.32\% with 40 samples, demonstrating the model's learning capability with more data. However, rate improvement slows down beyond 40 samples, only reaching 98.68\% with 80 samples, indicating a learning saturation. Still, with 100 samples, the rate peaks at 98.89\%, showing steady improvement. This analysis confirms the robustness of our method, even with limited samples, for SAR ATR.

Table \ref{comparisonMSTAR} is the comparison with other state-of-the-art methods for SAR ATR with limited data. Comparatively, the recognition performance significantly decreases when the per-class image number reduces from all data to 20 samples. MGAN-CNN mildly improves performance under 20 and 40 samples, while Semisupervised enhances it under 20, 40, and 80 samples, utilizing self-consistent augmentation and training resources. Quantitatively, our CIATR outperforms others under any number of training images per class, particularly under limited SAR training samples.

From the recognition performances and comparisons above, the superiority and effectiveness of our method have been validated.

\begin{table}[tb]
\renewcommand{\arraystretch}{1.2}
\setlength\tabcolsep{2.0pt}
\centering
\footnotesize
\caption{Recognition Performance (\%) under SOC on MSTAR}
\label{soc}
\begin{tabular}{c|cccccccc}
\toprule \toprule 
\multirow{2}{*}{Class} & \multicolumn{8}{c}{Training Number in   Each Class}                                  \\ \cline{2-9} 
                      & 5     & 10      & 25    & 30    & 40    & 60    & 80    & 100   \\ \hline
BMP2-9563                                       & 47.18               & 66.15                & 80.00  & 90.77  & 87.69  & 93.33  & 94.36  & 95.90  \\
BRDM2-E71                                      & 88.32               & 94.16                & 98.18  & 98.18  & 99.27  & 100.00 & 100.00 & 99.64  \\
BTR60-7532                                      & 62.05               & 58.97                & 95.90  & 93.33  & 95.38  & 96.92  & 98.46  & 97.44  \\
BTR70-c71                                      & 59.69               & 81.63               & 92.86  & 92.86  & 92.35  & 96.94  & 96.94 & 95.92  \\
D7-92                                         & 55.11               & 92.70                & 100.00 & 99.27  & 100.00 & 99.64  & 99.64  & 100.00 \\
2S1-b01                                        & 82.85               & 83.58               & 96.72  & 97.45  & 96.35  & 96.72  & 96.72  & 98.91  \\
T62-A51                                        & 89.38               & 87.91                & 99.27  & 98.53  & 99.63  & 99.27  & 100.00 & 99.63  \\
T72-132                                        & 69.39               & 92.35                & 98.47  & 98.98  & 97.96  & 100.00 & 98.98  & 99.49  \\
ZIL131-E12                                     & 93.43               & 93.80                & 100.00 & 99.27  & 100.00 & 100.00 & 100.00 & 100.00 \\
ZSU234-d08                                    & 85.40               & 100.00              & 100.00 & 100.00 & 100.00 & 100.00 & 100.00 & 100.00 \\
\midrule 
Average                                    & 75.05              & 86.47                & 96.70  & 97.24  & 97.32 & 98.47  & 98.68  & 98.89  \\ \bottomrule \bottomrule 
\end{tabular}
\end{table}

\begin{table}[tb]
\renewcommand{\arraystretch}{1.2}
\setlength\tabcolsep{9.5pt}
\centering
\footnotesize 
\caption{Comparison of Performance (\%) under SOC of MSTAR.}
\label{comparisonMSTAR}
\begin{tabular}{lcccc}
\toprule\toprule
\multirow{2}{*}{Algorithms} & \multicolumn{4}{c}{Image Number for Each Class}
\\ \cline{2-5} & 20    & 40    & 80    & All data  \\ \midrule
PCA+SVM \cite{comparison1}               & 76.43             & 87.95             & 92.48             & 94.32                 \\
ADaboost \cite{comparison1}              & 75.68             & 86.45             & 91.45             & 93.51                 \\
LC-KSVD \cite{comparison1}               & 78.83             & 87.39             & 93.23             & 95.13                 \\
DGM \cite{comparison1}                   & 81.11             & 88.14             & 92.85             & 96.07                 \\
DNN1 \cite{comparison2}                  & 77.86             & 86.98             & 93.04             & 95.54                 \\
DNN2 \cite{comparison3}                  & 79.39             & 87.73             & 93.76             & 96.50                 \\
CNN1 \cite{comparison1}                  & 81.80             & 88.35             & 93.88             & 97.03                 \\
CNN2 \cite{comparison2}                  & 75.88             & -                 & -                 & -                     \\
CNN+matrix \cite{comparison2}            & 82.29             & -                 & -                 & -                     \\
GAN-CNN \cite{comparison1}               & 84.39             & 90.13             & 94.91             & 97.53                 \\
MGAN-CNN \cite{comparison1}              & 85.23             & 90.82             & 94.91             & 97.81                 \\
Semisupervised \cite{open4}   & 92.62             & 97.11             & 98.65             & -                     \\
Ours                                     & -    & \textbf{97.32}    & \textbf{98.68}    & -                     \\
\bottomrule\bottomrule
\end{tabular}
\end{table}

\section{Conclusion}
In conclusion, we have proposed the causal interventional ATR method (CIATR) to address the challenges of SAR ATR with limited training data. Our approach leverages causal inference and backdoor adjustments to mitigate the effects of varying imaging conditions.
The structural causal model (SCM) helps us understand the role of imaging conditions as confounders in introducing spurious correlations in ATR models when data is limited. By using data augmentation with spatial-frequency domain hybrid transformation and feature discrimination with hybrid similarity measurement, we effectively estimate and mitigate the impacts of imaging conditions on SAR image features.
The CIATR method enables us to establish causal relationships between SAR images and their corresponding classes, even with limited data. Experimental results on the MSTAR and OpenSARship datasets validate the effectiveness of our approach in improving SAR ATR performance under limited training conditions.

\bibliographystyle{IEEEtran}
\bibliography{reference, causal, mypub}

\end{document}